\documentclass[11pt]{article}
\usepackage{amssymb,psfig}

   \csname @addtoreset\endcsname{equation}{section}
    \csname @addtoreset\endcsname{figure}{section}
    \csname @addtoreset\endcsname{table}{section}

\newcounter{thanksnum}
\def\thanksnumber#1
{\setcounter{thanksnum}{\value{footnote}}\setcounter{footnote}{#1}%
                     \addtocounter{footnote}{-1}\footnotemark
                     \setcounter{footnote}{\value{thanksnum}}}
\def\newtheoremz#1{\@ifnextchar[{\@othmz{#1}}{\@nthmz{#1}}}

\def\@nthmz#1#2{%
\@ifnextchar[{\@xnthmz{#1}{#2}}{\@ynthmz{#1}{#2}}}

\def\@xnthmz#1#2[#3]{\expandafter\@ifdefinable\csname #1\endcsname
{\@definecounter{#1}\@addtoreset{#1}{#3}%
\expandafter\xdef\csname the#1\endcsname{\expandafter\noexpand
  \csname the#3\endcsname \@thmcountersepz \@thmcounterz{#1}}%
\global\@namedef{#1}{\@thmz{#1}{#2}}\global\@namedef{end#1}{\@endtheoremz}}}

\def\@ynthmz#1#2{\expandafter\@ifdefinable\csname #1\endcsname
{\@definecounter{#1}%
\expandafter\xdef\csname the#1\endcsname{\@thmcounterz{#1}}%
\global\@namedef{#1}{\@thm{#1}{#2}}\global\@namedef{end#1}{\@endtheoremz}}}

\def\@othmz#1[#2]#3{\expandafter\@ifdefinable\csname #1\endcsname
  {\global\@namedef{the#1}{\@nameuse{the#2}}%
\global\@namedef{#1}{\@thmz{#2}{#3}}%
\global\@namedef{end#1}{\@endtheoremz}}}

\def\@thmz#1#2{\refstepcounter
    {#1}\@ifnextchar[{\@ythmz{#1}{#2}}{\@xthmz{#1}{#2}}}

\def\@xthmz#1#2{\@begintheoremz{#2}{\csname the#1\endcsname}\ignorespaces}
\def\@ythmz#1#2[#3]{\@opargbegintheoremz{#2}{\csname
       the#1\endcsname}{#3}\ignorespaces}

\def\@thmcounterz#1{\noexpand\arabic{#1}}
\def\@thmcountersepz{.}
\def\@begintheoremz#1#2{ \trivlist \item[\hskip \labelsep{\bf #1\ #2}]}
\def\@opargbegintheoremz#1#2#3{ \trivlist
      \item[\hskip \labelsep{\bf #1\ #2\ (#3)}]}
\def\@endtheoremz{\endtrivlist}

\newtheorem{lemma}{Lemma}[section]

\newtheorem{corollary}{Corollary}[section]

\newtheorem{definition}{Definition}[section]
\newtheorem{remark}{Remark}[section]

\def\e{\varepsilon}

\def\defi{\stackrel{{\scriptscriptstyle \Delta}}{=}}

\def\imp{{\scriptscriptstyle imp}}

\def\d{\delta}

\def\O{\Omega}

\def\F{{\cal F}}
\def\w{\widehat}

\def\R{{\bf R}}
\def\E{{\bf E}}
\def\P{{\bf P}}

\def\s{\delta}

\def\ww{\widetilde}

\def\s{\sigma}

\def\p{\partial}

\newcommand{\be}{\begin{equation}}
\newcommand{\ee}{\end{equation}}
\newcommand{\bd}{\begin{displaymath}}
\newcommand{\ed}{\end{displaymath}}
\newcommand{\ba}{\begin{array}{ll}}
\newcommand{\ea}{\end{array}}
\newcommand{\baa}{\begin{eqnarray}}
\newcommand{\eaa}{\end{eqnarray}}
\newcommand{\baaa}{\begin{eqnarray*}}
\newcommand{\eaaa}{\end{eqnarray*}}   \font\sm=cmr10

\def\BS{{\scriptscriptstyle BS}}
\def\RN{{\scriptscriptstyle RN}}

\def\ww{\tilde}

\def\QQ{{ Q}}
\date{ }
\title{Two unconditionally implied parameters and volatility smiles and
skews\footnote{Web-published: May 25, 2004 at
http://ssrn.com/abstract=550983. Published: {\it Applied Financial
Economics Letters} 2006, {\bf 2}  199-204. DOI:
10.1080/17446540500426771. Updated: April 22, 2013. Sections
\ref{SecE}-\ref{SecG} were added in 2013; they were not included in
the printed version.}}
\author{
Nikolai Dokuchaev
\thanks{Department of Mathematics and Statistics, University of
Limerick, Ireland and Department of Mathematics \& Statistics, Curtin
University,  GPO Box U1987, Perth, 6845 Western Australia}}
\begin{document}
\maketitle
\begin{abstract}
The paper studies estimation of parameters of diffusion market
models from historical data.  The standard definition of implied
volatility for these models presents its value as an implicit
function of several parameters, including the risk-free interest
rate. In reality, the risk free interest rate is unknown and need
to be forecasted, because the option price depends on its future
curve. Therefore, the standard implied volatility is  {\it
conditional}: it depends on the future values of the risk free
rate. We study two implied parameters: the implied volatility and
the implied average cumulative risk free interest rate. They can
be found unconditionally from  a system of two equations. We found
that very simple models with random volatilities (for instance,
with two point distributions) generate various volatility smiles
and skews with this approach.
\\ {\bf Key words}: market models, parameters estimation,
Black-Scholes,  implied volatility, implied forward risk-free rate, volatility
smile, volatility skew
\end{abstract}
{\it Short running head}: {\sm Two unconditionally implied
parameters}
\section*{Introduction}
 Most practitioners have adapted the famous
Black-Scholes model as the premier model for pricing and hedging
of options. This model consists of two assets: the risk free bond
or bank account and the risky stock. It is assumed that the
dynamics of the stock is given by a random process with some
standard deviation of the stock returns (the volatility
coefficient, or volatility).  Empirical research shows that the
real volatility is time-varying and random.  Many authors
emphasize that the main difficulty in modifying the Black--Scholes
and Merton models is taking into account  this fact.  A number of
equations  for evolution of the volatility were proposed (see e.g.
Christie (1982), Johnson and Shanno (1987), Hull  and  White
(1987), Masi {\it et al.} (1994), and more recent papers in Jarrow
(ed.) (1998)).   The basic pricing rule for models with random
volatility is risk neutral valuation, when the option price is
given as the expected value of its future payoff with respect to a
risk-neutral measure discounted back to the present time $t$ (see,
e.g., Ross (1976) and Cox and Ross (1976)). This method has been
developed to pricing rules based on optimal choice of the
risk-neutral measures such as local risk minimization, mean
variance hedging, $q$-optimal measures, and minimal entropy
measures
 (see, e.g.,
F\"ollmer and Sondermann (1986), Schweizer (1992),  Masi {\it et
al.} (1994), Geman {\it et al.} (1995), Rheinl\"ander and
Schweizer (1997), Pham {\it et al.} (1998), Laurent and Pham
(1999), Frittelli (2000), and others). These pricing rules  are
applicable for the most complicated models with It\^o's equation
for volatility.
\par
 In reality, practitioners prefer to describe market imperfection
and deviations from log-normal Black-Scholes model in the terms of
the so-called {\it volatility smile} or {\it volatility skew}  for
the implied volatility. It is a certain shape  of the implied
volatility on $K$ given $S(0)$, where $K$ is the strike price,
$S(0)$ is the stock price; $\cup$-shape is usually referred as the
volatility smile, $\cap$-shape and others are referred as the
volatility skew. It is commonly recognized that Black-Scholes
formula gives unbiased estimation for at-money options only, and
it gives a systematic error for in-money and out-of-money options.
That means that  there is a gap between historical and implied
volatility that  generates  volatility smile or  skew (see,  e.g.
Black and Scholes (1972), Day and Levis (1992), Derman {\it et
al.} (1996), Hauser and Lauterbach (1997), Taylor and  Xu (1994).
A detailed review
 can be found in Mayhew (1995)).
 Therefore, there is a demand for models  consisting of
stock prices, option prices, and volatilities, that can cover
different shapes of volatility smiles and skews.
For instance, the risk
neutral valuation method generates volatility smiles
rather than skews.
\par
In the present paper, we found a  very simple model with random
volatilities and risk free rates (for instance, with two point
distributions) that generates various volatility smiles and skews.
Our approach can be described as the following. The standard
implied volatility definition gives its value as a function of the
risk-free interest rate $r$, the option price, the strike price,
the current stock price, and terminal time $T$. The standard
definition of the implied volatility ignores the fact that, in
reality, $r$ is unknown and need to be forecasted, because the
option price depends on its future (forward) curve. Therefore, the
standard implied volatility at time $t$ is a {\it conditional} one
and it depends on  the future curve $r(s)|_{s\in[t,T]}$. In fact,
the Black-Scholes price at time $t$ depends only the volatility
process and on $\rho(t)=(T-t)^{-1}\int_t^Tr(s)ds$, or on a single
parameter of this curve (see Lemma \ref{lemma1} below), even if
$r(\cdot)$ is random and depends on $(S,\s,w)$, where $w$ s the
driving Wiener process. We suggest to calculate the pair
$(\s_{imp}(t),\rho_{imp}(t))$ of two {\it unconditionally} implied
parameters, where $\s_{imp}(t)$ is the unconditionally implied
volatility, and $\rho_{imp}(t)$ is the unconditionally implied
value of $\rho(t)$. This pair can be found from a system of two
equations with option prices for different strike prices. Note
that the case when two parameters are inferred from option
historical prices has been addressed by several authors  but in
different setting (see, e.g., survey of Garcia {\it et al}
(2004)). Butler and Schachter (1996) suggested to use two call
options with different strike prices for calculation of implied
volatility distributions for the case of option prices obtained
via {the unbiased estimate} of option price for random volatility.
The mentioned paper addressed the case of implied risk-free rate,
but it was focused on the case of the implied stock prices and
volatility.
\par
Our main goal is a model for volatility skews and smiles. Using
numerical simulation, we show that even simplest models with the
random volatility and the risk free rate with two point
distributions generate various volatility smiles and skews.
 \section{Definitions} We consider the diffusion model of a
securities market  consisting of a risk free bond or bank account
with the price $B(t), $ ${t\ge 0}$, and
 a risky stock with price $S(t)$, ${t\ge 0}$. The prices of the stocks evolves as
 \be \label{S}
dS(t)=S(t)\left(a(t)dt+\s(t) dw(t)\right), \quad t>0, \ee where
$w(t)$ is a Wiener process,  $a(t)$ is an appreciation rate,
$\s(t)$ is a random  volatility coefficient.  The initial price
$S(0)>0$ is a  given deterministic constant.  The price of the
bond evolves as
\begin{equation}
\label{B}
B(t)=\exp\biggl(\int_0^{t}r(s)ds\biggr)B(0),
\end{equation}
where $r(t)\ge 0$ is a random process and $B(0)$ is given. \par
\par We  assume that $w(\cdot)$ is a
standard Wiener process on  a given standard probability space
$(\O,\F,\P)$, where $\O$ is a set of elementary events, $\F$ is a
complete $\s$-algebra of events, and $\P$ is a probability
measure.
\par
 Let $\F_t$
be a filtration generated by the currently observable data.  We
assume that the process  $(S(t),\s(t))$ is $\F_t$-adapted and that
$\F_t$ does not depend on  $\{w(t_2)-w(t_1)\}_{t_2\ge t_2\ge t}$.
In particular, this means that the process $(S(t),\s(t))$ is
currently observable and $\s(t)$ does not depend on
$\{w(t_2)-w(t_1)\}_{t_2\ge t_2\ge t}$. We assume that $\F_0$ is
the $P$-augmentation of the set $\{\emptyset,\O\}$, and that
$a(t)$ does not depend on $\{w(t_2)-w(t_1)\}_{t_2\ge t_2\ge t}$.
For simplicity, we assume that $a(t)$ is a bounded process.
\subsection*{Black-Scholes price}
Let $K>0$ be given. We shall consider two types of options:
vanilla call and vanilla put, with payoff function
$f(S(T))=F(S(T),K)$, where $F(S(T),K)=(S(T)-K)^+$ or
$F(S(T),K)=(K-S(T))^+$,  respectively. Here $K$ is the strike
price.
\par
 Let $T>0$ be fixed. Let $H_{\BS,c}(t,x,\s,r,K)$ and
$H_{\BS,p}(t,x,\s,r,K)$ denotes Black-Scholes prices  for the
vanilla put and call options with the payoff functions $F(S(T),K)$
described above under the assumption that $S(t)=x$,
$(\s(s),r(s))=(\s,r)$ $(\forall s>t)$, where $\s\in (0,+\infty)$
is non-random.  The Black-Scholes formula for call  can be
rewritten as
 \baa\label{BS}
 H_{\BS,c}(t,x,\s,r,K)=x\Phi(d_+(t,x,\s,r,K))-Ke^{-r(T-t)}\Phi(d_-(t,x,\s,r,K)), \\
H_{\BS,p}(t,x,\s,r,K)=  H_{\BS,c}(t,x,\s,r,K)-x+ Ke^{-r(T-t)},
\nonumber \eaa
  where
$$
 \Phi(x)
\defi\frac{1}{\sqrt{2\pi}}\int_{-\infty}^x
e^{-\frac{s^2}{2}}ds,
$$
  and where
\baa
  d_+(x,t,\s,r,K)&\defi&\frac{\log{(x/K)}+(T-t)r}{\s\sqrt{(T-t)}}+ \frac{\s\sqrt{(T-t)}}{2}\nonumber,\\
    d_-(x,t,\s,r,K)&\defi& d_+(x,t,\s,r,K)-\s\sqrt{(T-t)}.
    \label{dpm} \eaa
Set $$ \ww S(t)\defi S(t)\exp\biggl(-\int_0^tr(s)ds\biggr).
$$
\par We assume that there exist
 a risk-neutral measure $Q$ such that the process
 $\ww S(t)$ is a martingale under $Q$, i.e.,
  $\E_Q\{\ww S(T)\,|\F_t\}=\ww S(t)$, where
 $\E_{Q}$ is
 the corresponding expectation.
\par
For brevity, we shall denote by $H_{\BS}$ the corresponding
Black-Scholes prices different options, i.e., $H_{\BS}=H_{\BS,c}$
or $H_{\BS}=H_{\BS,p}$, for vanilla call, vanilla put
respectively. Let $$ v(t)\defi\frac{1}{T-t}\int_t^T\s(s)^2ds,\quad
\rho(t)\defi \frac{1}{T-t}\int_t^Tr(s)ds. $$
\par
The following lemma is a generalization  for random $r(\cdot)$ of
the lemma from Hull and White (1987), p.245.
\begin{lemma}\label{lemma1}  Let $t\in[0,T)$ be fixed. Let
$v(t)$ and
 $\rho(t)$  be  $\F_t$-measurable.
Then $$
\E_{\QQ}\{e^{-\int_t^Tr(s)ds}F(S(T))|\F_t\}=H_{\BS}(t,S(t),\sqrt{v(t)},\rho(t),K).
$$
\end{lemma}
\par
Clearly, $\frac{1}{T-t}\int_t^T\s(s)^2ds$ and
$\frac{1}{T-t}\int_t^Tr(s)ds$ are not $\F_t$-measurable in the
general case of stochastic $(r,\s)$,  and the assumptions of Lemma
\ref{lemma1} are not satisfied.
\par
{\it Proof of Lemma \ref{lemma1}.}  It suffices to consider the
case when $t=0$ and $v(0)$ and $\rho(0)$ are non-random.
\par Set   $ \ww K\defi
K\exp\biggl(-\int_0^Tr(s)ds\biggr). $ We introduce the function
$\w H_\BS(\cdot):[0,T]\times\R\to\R$ such that
$$e^{\rho(0) t}\w H_{\BS}(t,x)\equiv H_{\BS}(t,e^{\rho(0)
t}x,\sqrt{v(0)},\rho(0),K).$$ It is easy to see that $$\ba
\frac{\p\w H_{\BS}}{\p t}(t,x)+\frac{1}{2}v(0)x^2\frac{\p^2\w
H_{\BS}}{\p x^2}(t,x)=0,
\\\w H_{\BS}(T,x)=F(x,\ww K). \ea $$
Let $$ \tau(t)\defi \frac{1}{v(0)}\int_0^t\s(s)^2ds,\quad\ww
X(t)\defi\ww H_{\BS}(\tau(t),\ww S(t)), \quad \ww
S(t)\defi\exp\left(\int_0^tr(s)ds\right)S(t). $$ By Ito formula,
we obtain that
$$ d\ww X(t)=\frac{\p \w H_{\BS}}{\p x}(\tau(t),\ww S(t))d\ww
S(t), \quad \ww X(T)=F(\ww S(T),\ww K). $$ Hence $$ \ww X(0)=\w
H_{\BS}(0,S(0))=\E_{Q} F(\ww S(T),\ww K
)=\E_{Q}\exp\left(-\int_0^Tr(s)ds\right)F(S(T),K).$$ This
completes the proof.
 $\Box$
 \begin{corollary}\label{lemmBS}
 Assume that   $H_{\BS}=H_{\BS,c}$, or $H_{\BS}=H_{\BS,p}$,  or  $H_{\BS}=H_{\BS,s}$.
 Consider a market model with pricing rule (\ref{rule}).
 Let $(\s,r)$ does not depend on $w$ under $\QQ$. Then
$P_{\RN}(t)=\E_Q\{
H_{\BS}(t,S(t),\sqrt{v(t)},\rho(t),K)\,|\,\F_t\}$, where
$(v,\rho)$ are defined in Lemma \ref{lemma1}.
\end{corollary}
 \subsection*{Unconditionally
implied parameters} The standard definition of the implied
volatility ignores the fact that, in reality, $r$ is unknown and
need to be forecasted, because the option price depends on its
future (forward) curve. Therefore, the standard implied volatility
at time $t$ is a {\it conditional} one and it depends on  the
future curve $r(s)|_{s\in[t,T]}$. We shall study the pair
$(\s_{imp}(t),\rho_{imp}(t))$ of two {\it unconditionally} implied
parameters, where $\s_{imp}(t)$ is the unconditionally implied
volatility, and $\rho_{imp}(t)$ is the unconditionally implied
value of $\rho(t)$. This pair of implied parameters can be
inferred from a system of two equations for different options.
\begin{definition}\label{def1} Assume that we observe
two options on the same stock
 with market prices $P^{(1)}(t)$ and $P^{(2)}(t)$ at time $t$.  These
options have the same expiration time $T>0$.  Let  $H_{\BS}^{(1)}$
and
 $H_{\BS}^{(2)}$ be the Black-Scholes price for the corresponding types of options. Let the pair
$(\s_{\imp}(t),\rho_{\imp}(t))$ be such that
\be
\left\{\ba H_{\BS}^{(1)}(t,S(t),\s_{\imp}(t),
\rho_{\imp}(t))=P^{(1)}(t),\\
 H_{\BS}^{(2)}(t,S(t),\s_{\imp}(t), \rho_{\imp}(t))=P^{(2)}(t).\ea\right.
\label{system0} \ee
 We say that
$\s_{\imp}(t)$ is the {\it implied volatility} and
$\rho_{\imp}(t)$ is the {\it implied average forward risk-free
rate} inferred from (\ref{system0}).
\end{definition}
To avoid technical difficulties, we shall assume that the prices
and parameters in (\ref{system0}) are such that the solution
$(\s_{\imp}(t),\rho_{\imp}(t))$ exists and is uniquely defined for
all special case described below.
 Clearly, Definition \ref{def1} is
model free and does not require any pricing rules and a prior
assumptions on the evolution law for volatilities and risk free
rates. We need some models and pricing rules only for numerical
simulations of
  of $(\s_{\imp}(t),\rho_{\imp}(t))$.
\subsubsection*{Pricing rule} The local risk minimization method,
the mean variance hedging, and some other
 methods based on the risk-neutral valuation lead
to the following pricing rule: given $(a,\s,r)$, the option price
is \be\label{rule} P_{\RN}(t,\s(\cdot),r(\cdot))\defi
\E_{Q}\{e^{-\int_t^Tr(s)ds}F( S(T))\,|\,\F_t\}, \ee where $Q$ is
some risk neutral measure, and where $\E_{Q}$ is the corresponding
expectation. Usually, $Q$ is uniquely defined by $(a,\s,r)$, and
by the pricing method used.
\par For numerical simulation purposes, we assume that we have chosen one of these methods (for instance,
local risk minimization method or mean variance hedging).
Therefore, the risk neutral measure $Q$ is uniquely defined by
$(a,\s,r)$ given the method of pricing.
\section{Two
calls with different strike prices} Assume that two  European
call options on the same stock
 have
market prices $P_i(t)$ at time $t$, $i=1,2$. We assume that these
options have the same expiration time $T>0$ and have different
strike prices $K_i>0$,  $K_1\neq K_2$. Let
$\s_{\imp}(t)=\s_{\imp}(t,K_1,K_2)$ be the {\it implied
volatility} and $\rho_{\imp}(t)=\rho_{\imp}(t,K_1,K_2)$ be the
{\it implied average forward risk-free  rate} given $K_1,K_2$ at
time $t$, inferred from the system \be \left\{\ba
H_{\BS,c}(t,S(t),\s_{\imp}(t), \rho_{\imp}(t),K_1)=P_1(t),\\
 H_{\BS,c}(t,S(t),\s_{\imp}(t), \rho_{\imp}(t),K_2)=P_2(t).\ea\right.
\label{system} \ee
  \begin{remark}{\rm For solution of system (\ref{system}),
  the following straightforward algorithm can be
  applied.
 Let $\s_1(t,K_1|\,\rho)$
 be such that
$$ H_{\BS,c}(t,S(t),\s_{1}(t,K_1|\,\rho), \rho, K_1)=P_1(t). $$  (i.e.
it is the standard (conditional) implied volatility). Consider
equation \be\label{scalar}
H_{\BS,c}(t,S(t),\s_{1}(t,K_1|\,\rho),\rho, K_2)=P_2(t). \ee Let
$\w\rho=\w\rho(K_1,K_2)$ be the solution  of (\ref{scalar}).
Then
$$(\s_{\imp}(t),\rho_{\imp}(t))=\left(\s_{1}(t,K_1|\,\w\rho(K_1,K_2)),\w\rho(K_1,K_2)\right).$$
  }\end{remark}
\subsection*{Numerical simulation for generic market model}
For numerical simulation, we accept the simplest stock market
model with traded options on that stock and with pricing rule
(\ref{rule}). Assume that the risk neutral measure $Q$ is such
that the process $(r(t),\s(t))=(\s,r)$ is random, independent on
$w$ under $Q$, independent on time, and can take only two values,
$(r_1,\s_1)$ and $(r_2,\s_2)$, with probabilities $p$ and $1-p$
correspondingly, where $p\in [0,1]$ is given. In that case,
pricing rule (\ref{rule}) means that the price of call option with
strike price $K$ and expiration time $T$ is $$ \E_Q\max(0,\ww
S_T-\ww
K)=p\,H_{\BS}(T,S_0,r_1,\s_1,K)+(1-p)\,H_{\BS}(T,S_0,r_2,\s_2,K).
$$ Clearly,  any $p\in[0,1]$ defines its own risk-neutral
probability measure $Q$, and, therefore, it defines its own
$\E_Q$.
\par
\begin{figure}
\centerline{\psfig{figure=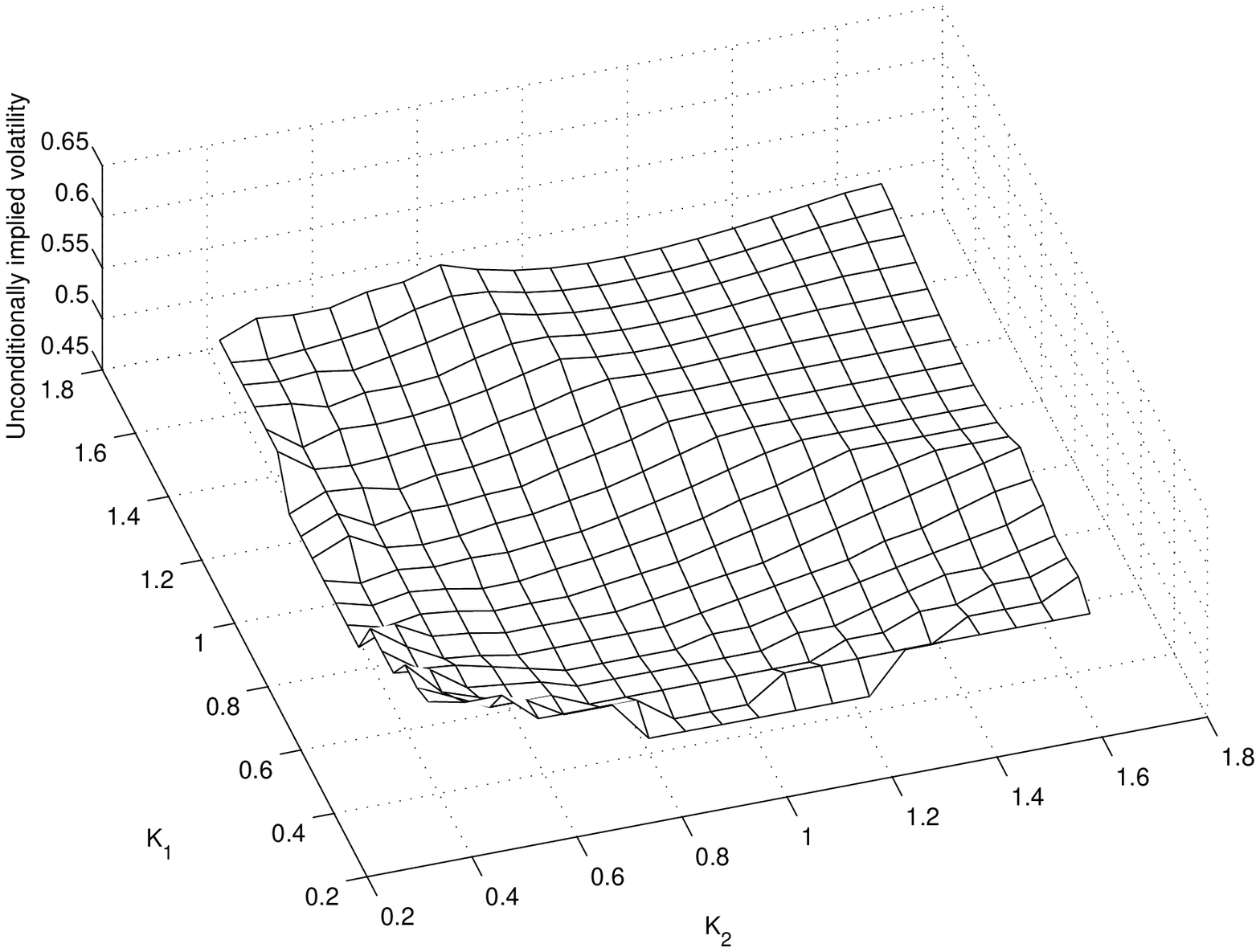,height=7cm}}
\centerline{\psfig{figure=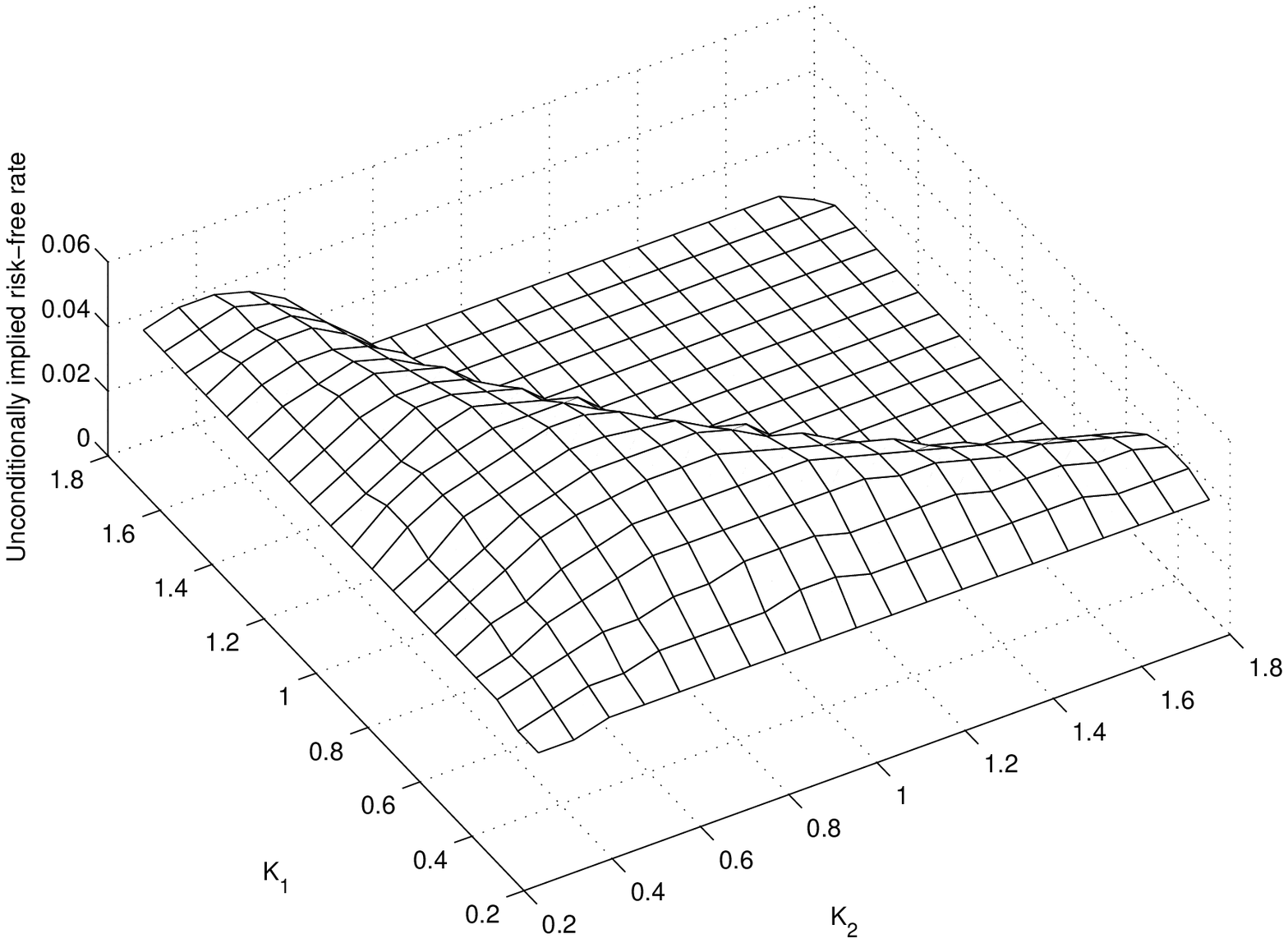,height=7cm}}
\par
\caption{{\sf Unconditionally implied volatility
$\s_{\imp}(t,K_1,K_2)$    (top)
and average forward risk-free rate (bottom)
 $r_{\imp}(t,K_1,K_2)$ inferred from prices for two call options
for the case  when  $t=0$, $T=1$, $S(0)=1$, $p=0.5$, $\s_1=0.3$,
$\s_2=0.7$, $r_1=0.01$, $r_2=0.08$. }} \label{fig1}
\end{figure}

As an example, we consider the case when  $t=0$, $T=1$, $S(0)=1$,
$p=0.5$, $\s_1=0.3$, $\s_2=0.7$, $r_1=0.1$, $r_2=0.08$. Figure
\ref{fig1} shows the unconditionally implied volatility and
average forward risk-free rate
 $(\s_{\imp}(t,K_1,K_2),r_{\imp}(t,K_1,K_2))$
\par
\begin{figure}
\centerline{\psfig{figure=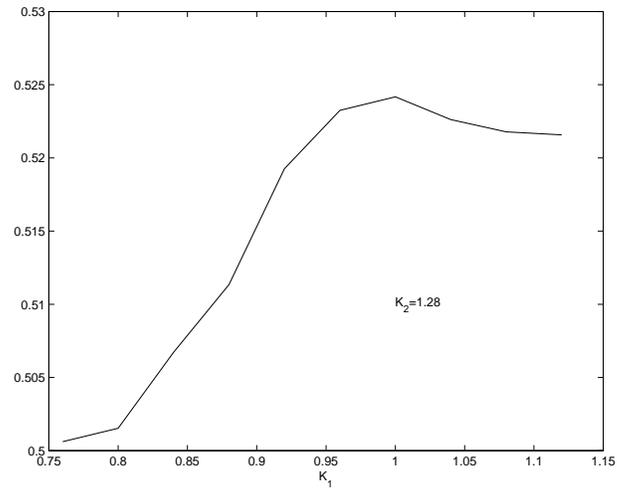,height=7cm}}
\centerline{\psfig{figure=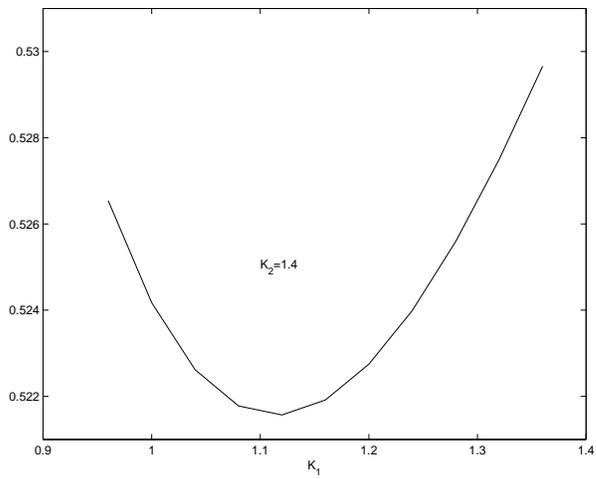,height=7cm}}
\par
\caption{{\sf Unconditionally implied volatility
$\s_{\imp}(t,K_1,K_2)$ inferred from prices for two call options
given $K_2=1.28$ and $K_2=1.4$.}} \label{fig2}
\end{figure}

 Figure \ref{fig2} shows the shape of dependence of   unconditionally implied volatility
$\s_{\imp}(t,K_1,K_2)$  on $K_1$ given  $K_2=1.28$ and $K_2=1.4$.
\par
It can be seen that the volatility surface the risk-free rate
surface neither convex nor concave  with respect to $(K_1,K_2)$,
and our simplest model can generate volatility smiles as well as
skews.
\section{Exclusion of  the stock prices form the system of equation}
\label{SecE} For the
dynamic estimation of time varying  implied parameters, one has to separate the impact of the changes of the stock price on the option price   from the impact of the change of the values of the
stock price parameters. For this, it could be convenient  to
exclude the current stock price form the system of equations for the implied parameters. To address it, we suggest the following approach.

Let us consider dynamically adjusted  parameters  $T=t+\tau$ and $K=\kappa S(t)$, where
$\kappa\in(0,+\infty)$ is a parameter. In this case,
$F(S(T))=F(S(T),K)=S(t)F(Y(t+\tau),\kappa)$, where
$$Y(T)=S(t+\tau)/S(t).$$ By rule (\ref{rule}), the option price  given
$(a,\s,r)$, is \baaa\label{rule1}
P_{\RN}(t,\s(\cdot),r(\cdot))&\defi& \E_{Q}\{e^{-\int_t^Tr(s)ds}F(
S(T),K)\,|\,\F_t\}\nonumber\\&=&
S(t)\E_{Q}\{e^{-\int_t^{t+\tau}r(s)ds}F(
Y(t+\tau),\kappa)\,|\,\F_t\}, \eaaa where $Q$ is some risk neutral
measure, and where $\E_{Q}$ is the corresponding expectation.

Let \baaa
G(t)\defi \frac{P_{\RN}(t,\s(\cdot),r(\cdot))}{S(t)}.\eaaa It follows that
\baaa\label{rule2} G(t)=\E_{Q}\{e^{-\int_t^{t+\tau}r(s)ds}F(
Y(t+\tau),\kappa)\,|\,\F_t\}, \eaaa
Therefore, the implied parameters  $(\s_{\imp}(t), \rho_{\imp}(t))$ for European options can be calculated using $H_{\BS}(t,1,\s,\rho,\d)
$ only with $\kappa=\kappa_i$, $i=1,2$, $\kappa_1\neq \kappa_2$. 

Let us consider the following example. Assume that two  European call options on the
same stock
 have
market prices $P_i(t)$ at time $t$, $i=1,2$.  We assume that these options have the same
expiration time $T=t+\tau>0$ and have different strike prices
$K_i=\kappa_i S(t)>0$,  $\kappa_1\neq \kappa_2$. Let
$G_i(t)=P_i(t)/S(t)$.  In this case,  the { implied volatility}   $\s_{\imp}(t)$ and  the {\ implied
average forward risk-free rate} $\rho_{\imp}(t)$ at time $t$ can be 
inferred from the system \be \left\{\ba
H_{\BS,c}(t,1,\s_{\imp}(t), \rho_{\imp}(t),\kappa_1)=G_1(t),\\
 H_{\BS,c}(t,1,\s_{\imp}(t), \rho_{\imp}(t),\kappa_2)=G_2(t).\ea\right.
\label{system} \ee
\begin{remark}{\rm The observations of option prices with  dynamic adjusted strike price $K=\kappa S(t)$ with a fixed $\kappa$ can be 
 useful for econometrics purposes even without calculation of the implied parameters. In particular, some features of the evolution law
for historical parameters  $(\s(t),\rho(t))$ can be restored directly from the observations of the process  $G(t)$. For instance, the processes $G(t)$ must evolve as
a deterministic function of the current values of $(\s(t),\rho(t))$ if the process  $(\s(t),\rho(t))$ evolves as a Markov process  that
is independent from $w(\cdot)$.}\end{remark} \footnote{Section \ref{SecE} was not
included in the printed version; it was added in the web-published version
on April 22, 2013.}
\section{Possible generalizations }\label{SecG}
The approach suggested in this paper allows many straightforward generalizations.  For instance,
assume that implied parameters are calculated using market prices of three options  with expiration times $T_1,T_2, T_3$ such that $T_1<T_2=T_3$.
Let $\s_1,\s_2,\s_3,\rho$ be the corresponding implied volatilities and the implied cumulative risk free rate calculated  as the
solution of the system of the three equations for prices; we  assume  that $\s_2=\s_3$. The relationship between $\s_1$ and $\s_2=\s_3$ shows the implied market hypothesis about the evolution  of the volatility.
\par
Furthermore,  sets of special implied  parameters  can be used for models that are different from the Black-Scholes diffusion market model.
For example, consider a model with driving fractional Brownian motion with unknown  Hurst parameter $h$. A system of three equations including the prices for three options
  can be used to determine the implied $(\rho,\s,h)$, where $\rho$ is the implied risk-free rate, $\s$ is the implied risk-free rate, $h$ is the implied Hurst parameter.
  Instead of the classical Black-Scholes formula for the  prices, one should use the corresponding modification of the pricing formula for the case of  fractional Brownian motion.
  \par
The same approach for can be applied for the discrete time market models.  Let us consider the so-called binomial model.
   Let us suggest an example of a pair of implied parameters associated with a modification of this model with the prices $S(t)$, $t=0,1,2,...,N$.
   We assume that the evolution of $S(t)$ is such that  $S(t+1)=\rho S(t)\zeta(t+1)$, $t=0,1,2,...$, where $\P(\zeta(t+1)\in\{1-\e,(1-\e)^{-1}\}|\F_t)=1$,  where  $\F_t$ is the filtration generated by $S(t)$, $\rho\ge 1$ is the single period return for the risk free investment, $\e\in (0,1)$ is a parameter for the model.
   Instead of the classical Black-Scholes formula for  the option price, one can use the value of  the initial wealth that allows replication of the claim $\rho^{-N}F(S(N))$. The pair  $(\rho,\e)$ can be used as the pair of implied parameters; $\rho$ represents the single period return for the risk-free investment, and $\e$ represents  the range of change that can be considered as
   an analog of the volatility.
\par
   For the classical binomial model, the sets of all possible values of the
  stock prices are finite at every time.
   Let us suggest  a modification of the discrete time  binomial model such that the distribution of the stock price is continuous and the  stock prices can take any positive value.
   Let us consider first a  model for "rounded" prices $S_0(t)$ with a finite set of possible values at any time $t=0,1,2,...$.
   We assume that the evolution of $S_0(t)$ is described by a standard discrete time binomial model such that $S_0(0)=1$,  $S_0(t+1)=\rho S(t)\zeta(t+1)$, $t=0,1,2,...$, where $\P(\zeta(t+1)\in\{1-\e,(1-\e)^{-1}\}|\F_t)=1$,   $\F_t$ is the filtration generated by $S_0(t)$, $\rho\ge 1$ is the single period return for the risk free investment, $\e\in (0,1)$ is a parameter for the model.
Second, let us consider a sequence of random variables $\xi(t,v)$ such that are mutually independent given $\F_t$ and that they all have the uniform distribution on $[(1-\e),(1-\e)^{-1}]$ conditionally given $\F_t$, where $v\in V_t$. Here $V_t$ is the supporting set for the distribution  of $S_0(t)$ . Finally, let us select the final model for the stock prices to be   $S(t+1)=\rho S(t)\zeta(t+1)\xi(t+1,S_0(t))$. For this model, $S(t)$ can take any positive value. The pair   $(\rho,\e)$ can be used as the pair of implied parameters again; instead of the Black-Scholes pricing formula, one can use the price calculated for the binomial stock price model described by $S_0(t)$.
\footnote{Section \ref{SecG} was not included in the printed version; it was added in the web-published version on March 20, 2013.}
\vspace{3mm}
\par
 The author wishes to thank Barry
 Schachter for useful comments regarding the bibliography.
\section*{ References}$\hphantom{xx}$ Black, F. and M. Scholes
(1972): The valuation of options contracts and test of market
efficiency. {\it Journal of Finance}, {27}, 399-417.
\par
Butler, J.S. and B. Schachter (1996): Statistical Properties of
Parameters Inferred from the Black-Scholes Formula August.
International Review of Financial Analysis {\bf 5}, 223-235.
\par Christie, A. (1982): The stochastic behavior of
common stocks variances: values, leverage, and interest rate
effects. {\it Journal of Financial Economics}, {10},
407-432.
\par Day, T.E. and C.M. Levis (1992): Stock
  market volatility and the information
content of stock index options. {\it Journal of Econometrics},
{52}, 267-287.
\par
Cox, J. C., and S.A. Ross (1976): The valuation of options for alternative stochastic
processes, {\it Journal of Financial Economics} {\bf 3}, 145–166.
\par
Derman, E., I. Kani,  and J.Z. Zou (1996): The local volatility
surface: unlocking the information in index option prices. {\it
Financial Analysts Journal} 25-36.\par Geman, H. and T. Ane
(1996): Stochastic subordination. {\it Risk}, {9} (9), 145-149.
\par
Fleming, W., and H.M. Soner  (1993): {\it Controlled Markov Processes and Viscosity Solutions.}
Applications
Math., vol. 25. Berlin-Heidelberg-New York: Springer.
\par
F\"ollmer, H., and D. Sonderman (1986): Hedging of Non-Redundant
Contingent Claims. In: Mas-Colell, A., Hildebrand, W. (eds.) {\it
Contributions to Mathematical Economics}. Amsterdam: North Holland
1986, pp. 205–223.
\par
Frittelli, M. (2000): The minimal entropy martingale measure and the
valuation problem in incomplete markets. {\it Mathematical Finance}, {\bf 10}, 39-
52.
\par
Geman H., N.El Karoui, J.-C.Rochet. (1995): Changes of numeraire,
changes of probability measure and option pricing. {\it Journal of
Applied Probability} {\bf 32} 443-458.
\par
 Garcia, R., Ghysels, E., Renault, E.
(2004): The Econometrics of option pricing, Working Paper CIRANO
2004s-04 (to appears in Handbook of Financial Econometrics, Y.
Ait-Sahalia and L.P. Hansen (eds.) North Holland).
\par Hauser, S.
and B. Lauterbach (1997): The relative performance of five
alternative warrant pricing models. {\it Financial Analysts
Journal},  N1, 55-61.
\par Hull, J. and A. White
(1987): The pricing of options on assets with stochastic
volatilities. {\it Journal of Finance}, { 42},
281-300.
\par Jarrow, R. (ed.)
(1998): {\it Volatility New Estimations Techniques for pricing
Derivatives}, Risk Books.
\par Johnson, H. and D. Shanno
(1987): Option pricing when the variance is changing. {\it Journal
of Financial and
Quantitative Analysis}, { 22}, 
143-151.                               \par
 Lambertone, D., and B. Lapeyre (1996): {\em Introduction
to Stochastic Calculus Applied to Finance.} London: Chapman \&
Hall.
\par
Laurent, J.P., and H. Pham  (1999): Dynamic programming and
mean-variance hedging. {\it Finance and Stochastics} 3, 83–110.
\par
  Masi, G.B., Kabanov, Yu.M.,  and W.J. Runggaldier (1994):
 Mean-variance hedging of options on stocks
with Markov volatilities. {\it Theory of Probability and Its
Applications
} { 39}, 
172-182.
\par Mayhew, S. (1995): Implied volatility. {\it Financial
Analysts Journal}, iss. 4,  8-20.
\par
 Pham, H., Rheinlander, T., and M. Schweizer (1998):
Mean-variance hedging for continuous processes: new proofs and
examples. {\em Finance and Stochastics} {\bf 2}, 173--198.
\par
Rheinl\"ander, T., and M. Schweizer (1997): On $L^2$-projections
on a space of stochastic integrals. {\it Annals of Probability}
{\bf 25}, 1810-1831.
\par
Ross, S. (1976): Options and efficiency,
{\it Quarterly Journal of Economics} {\bf  90}, 75–89.
\par
Schweizer, M. (1992): Mean-Variance Hedging for General Claims.
{\it Annals of Applied Probability} 2, 171-179.
\par
 Taylor, S.J. and  X. Xu (1994): The magnitude of
implied volatility smiles: theory and empirical evidence for
exchange rates.
{\it Review of Future Markets}, {\bf 13}, 
355-380.
               \end{document}